\documentclass[journal]{IEEEtran}
\IEEEoverridecommandlockouts
\usepackage{cite}

\usepackage{amsmath, amssymb, amsfonts, mathtools, bm}
\usepackage{amsthm}
\usepackage{algorithm}
\usepackage{algorithmic}
\usepackage{graphicx}
\usepackage{textcomp}
\usepackage{xcolor, float}
\usepackage{tabularx}
\usepackage{textcomp}
\usepackage{diagbox}
\usepackage{svg}
\usepackage{booktabs}
\usepackage{subfigure}
\usepackage{setspace}
\usepackage{multirow}
\usepackage{xcolor}
\usepackage{array}
\usepackage{color}
\usepackage{colortbl}  
\usepackage{xcolor}
\usepackage{array} 
\usepackage{makecell}
\DeclareUnicodeCharacter{2061}{}

\ifCLASSINFOpdf
\else
\fi
\begin{document}
%
\title{Knowledge-Driven Deep Learning Paradigms for Wireless Network Optimization in 6G}

%

\author{ Ruijin Sun, 
         Nan Cheng, 
         Changle Li,          
         Fangjiong Chen,
         Wen Chen 
\thanks{Corresponding author: Nan Cheng.}
\thanks{R. Sun, N. Cheng and C. Li are with the State Key  Lab of ISN and the School of Telecommunications Engineering, Xidian University, Xi’an 710071,  China (email:  sunruijin@xidian.edu.cn, nancheng@xidian.edu.cn, clli@mail.xidian.edu.cn)). }
\thanks{F. Chen is with the School of Electronic and Information Engineering, South China University of Technology, Guangzhou 510641, China (e-mail:eefjchen@scut.edu.cn).}
\thanks{W. Chen is with the Department of Electronics Engineering, Shanghai Jiao Tong University, Shanghai 200240, China (e-mail: wenchen@sjtu.edu.cn).}
}

\IEEEaftertitletext{\vspace{-1\baselineskip}}

\maketitle

\begin{abstract}
 In the sixth-generation (6G) networks,  newly emerging diversified services of massive users in dynamic network environments are required to be satisfied by multi-dimensional heterogeneous resources. The resulting large-scale complicated  network optimization problems are beyond the capability of model-based theoretical methods due to the overwhelming computational complexity and the long processing time. Although with fast online inference and universal approximation ability, 
 data-driven deep learning (DL) heavily relies on abundant training data and lacks interpretability. To address these issues, a new paradigm called knowledge-driven DL has emerged,  aiming to integrate proven domain knowledge into the construction of neural networks, thereby exploiting  the strengths of both methods. This article provides a systematic review of knowledge-driven DL in wireless networks. Specifically, a holistic framework of knowledge-driven DL in wireless networks is proposed, where knowledge sources, knowledge representation, knowledge integration and knowledge application are forming as a closed loop. Then, a detailed taxonomy of knowledge integration approaches, including knowledge-assisted, knowledge-fused, and knowledge-embedded DL, is presented.  Several open issues for future research are also discussed. The insights offered in this article provide a basic principle for the design of network optimization that incorporates communication-specific domain knowledge and DL, facilitating the  realization of intelligent 6G networks.

\end{abstract}



%
\IEEEpeerreviewmaketitle

\section{Introduction}

With the standardization and swift commercialization of the fifth-generation (5G) wireless communication systems, global researchers in both academia and industry have taken a leap into the future and started the planning of sixth-generation (6G) wireless communication networks. As anticipated by the International Telecommunication Union Radiocommunication sector (ITU-R), the forthcoming era of 6G is expected to see the flourishing of a multitude of newly emerging wireless applications, ranging from immersive communications to integrated sensing and communications \cite{ITU2023framework}. To accommodate such diverse services in dynamic wireless environments, 6G networks are predicted to deeply integrate and flexibility schedule multi-dimensional resources, including computing, storage, spectrum, power, time, and space. This results in large-scale complicated network optimization problems,  which are beyond the capability of traditional model-based algorithms due to their prohibitive computational complexity and extensive online processing time. Consequently, the development of innovative intelligent resource scheduling strategies becomes an essential requirement for future 6G networks, \textcolor{black}{which is also in line with one of the 6G usage scenarios envisioned by ITU-R, i.e., artificial intelligence (AI) and communication \cite{ITU2023framework}.}

As a prominent branch of AI, deep learning (DL), also known as neural networks, has garnered considerable attention in the field of wireless communication networks, due to its unprecedented success in natural language processing and computer vision. Thanks to their universal approximation ability, neural networks have shown competitive performance in physical layer communications, network resource allocations, and network operation and management. However, most deep learning approaches are primarily data-driven, learning the complex mapping relationship based on the statistical distribution of data. This methodology requires a large amount of high-quality data, which is often scarce in wireless communication networks. Moreover, the internal mechanisms of neural networks remain elusive, leading to a perception of neural networks as ``black boxes'' that lack interpretability and reliability. To overcome these limitations, a new paradigm, knowledge-driven deep learning, has emerged, which seeks to construct neural networks inspired by proven domain knowledge accumulated over decades \cite{he2019modeldriven,shlezinger2021model}. \textcolor{black}{By investigating several examples of knowledge-driven DL for signal processing,} work in \cite{he2019modeldriven,shlezinger2021model} reveals that with the involvement of explainable knowledge, knowledge-driven DL methods have the potential to enhance the interpretability of neural networks, thereby reducing the required training data samples. Therefore, knowledge-driven DL, although in its infancy stage, is regarded as an appealing avenue for achieving reliable intelligent communications.

\textcolor{black}{To attain a comprehensive understanding of knowledge-driven DL for wireless network optimization, this article first proposes a holistic framework of knowledge-driven DL in wireless networks and systematically summarizes a  taxonomy of knowledge integration approaches.} The remainder of this article is organized as follows. The next section introduces the concept and advantages of knowledge-driven DL in wireless networks. Then, we present a holistic framework of knowledge-driven DL in network optimizations.  Following that we present a detailed taxonomy of knowledge integration approaches in wireless networks. We then discuss some open issues in this research area before the article is concluded.


\begin{figure*}[htbp]
\centering
\includegraphics[width=0.9\textwidth]{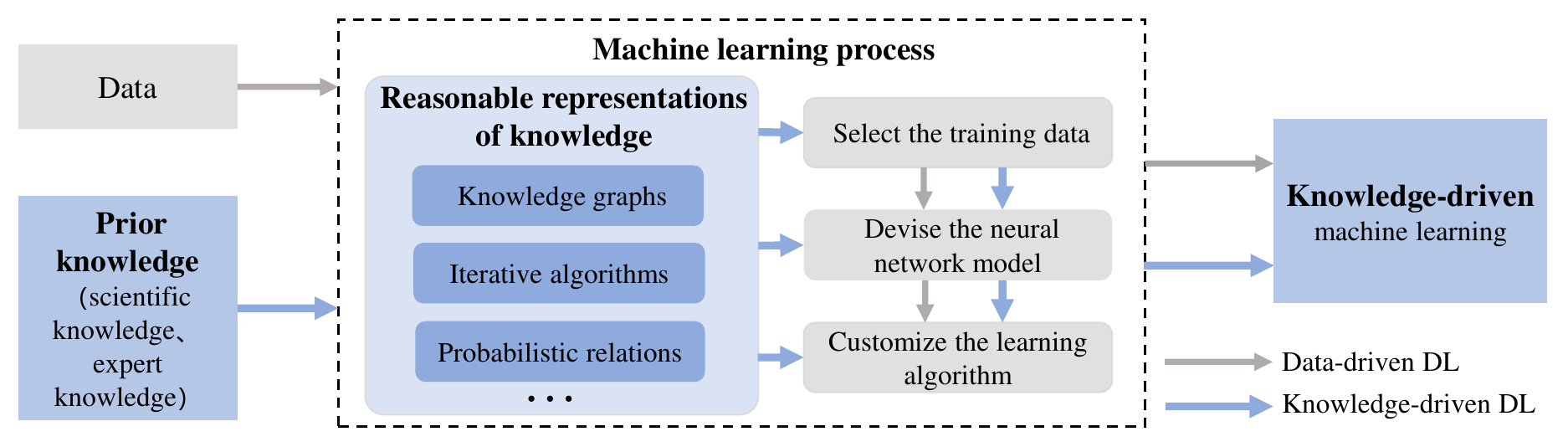}
\caption{The pipeline of knowledge-driven DL. Compared with data-driven DL, in knowledge-driven DL, prior knowledge with reasonable representations is integrated into neural networks to select the training data, devise the neural network model and customize the learning algorithms.}
\label{Fig.1}
\end{figure*}

\section{Why Knowledge-driven DL for 6G?}
In this section, we first introduce two existing optimization methods in wireless networks, i.e., model-based theoretical methods and data-driven DL methods, whose drawbacks motivate the development of knowledge-driven DL methods. Then, the concept of knowledge-driven DL methods is presented, along with their strengths in 6G.

\subsection{Existing Optimization Methods in Wireless Networks}
In wireless networks, existing optimization methods are mainly divided into two categories, i.e., model-based theoretical methods and data-driven DL methods. Model-based theoretical methods  refer to the techniques and algorithms used to optimize the wireless network totally depending on validated theoretical models, such as communication theory, signal processing, and optimization theory, which are termed as communication-specific domain knowledge. Relying on such domain knowledge, model-based methods optimize the wireless network with a series of explainable logic rules and mathematical formulas, allowing researchers and practitioners to understand the behavior of wireless networks.  However, 6G networks, with their diverse services, customized requirements, multi-dimensional resources, and large-scale connected nodes, pose significant challenges to model-based network optimization. First, to make complicated network optimization problems tractable, model-based methods usually make simplistic assumptions about  wireless networks, such as Gaussian noise and linear systems, leading to inaccurate system models and suboptimal solutions. Second, model-based methods suffer from overwhelming computational complexity in large-scale network optimization, resulting in prolonged inference time, which are unsuitable for services with stringent low-latency requirements.

Data-driven DL methods refer to the approaches that utilize neural networks to directly learn the network optimization policies from data. Specifically, network optimization problems are reformulated as regression tasks, where the network status (inputs) are mapped
to the policies  (outputs)  via neural networks and their parameters are learnt from the labeled dataset. As with universal approximation ability,  neural networks can map a wide range of relationships, allowing data-driven methods to handle complicated problems where analytical models may not be available. Another important advantage of neural networks is their fast  online inference although requiring  offline training, contributing to the quick response to dynamic networks. Nevertheless,  learning highly-parameterized neural networks requires a large amount of high-quality training data, whose collection is time-consuming or even impossible in wireless networks. Besides, the training process of data-driven methods involves heavy computational burden, which  might be challenging for hardware-limited communication devices. Furthermore, neural networks are treated as  ``black boxes", as their inner workings and mechanisms are not easily understood. This absence of interpretability hinders data-driven methods from offering performance guarantees in  network optimizations.


\subsection{Knowledge-Driven DL Methods in Wireless Networks}

To  address the issues faced by existing methods, a novel paradigm called knowledge-driven DL is proposed  for network optimization in 6G, \textcolor{black}{which adopts interpretable domain knowledge to guide the design of neural networks}. By incorporating communication-specific domain knowledge into neural networks, knowledge-driven DL methods leverage the long-standing expertise developed over decades in wireless communications to customize the construction and training of neural networks for wireless tasks. Specifically, a typical knowledge-driven DL pipeline  consists of four main components, i.e., the training data, the neural network model, the learning algorithm and domain knowledge, as illustrated in Fig. 1.  Compared with pure data-driven DL, the additional domain knowledge, also referred to as prior knowledge, is added in the pipeline, which is represented in various forms, such as knowledge graphs, iterative algorithms, probabilistic relations and so on. When constructing a knowledge-driven DL for a particular task, corresponding domain knowledge is utilized to select the training data, devise the neural network model and customize the learning algorithm. 


\begin{figure*}[htbp]
\centering
\includegraphics[width=1.0\textwidth]{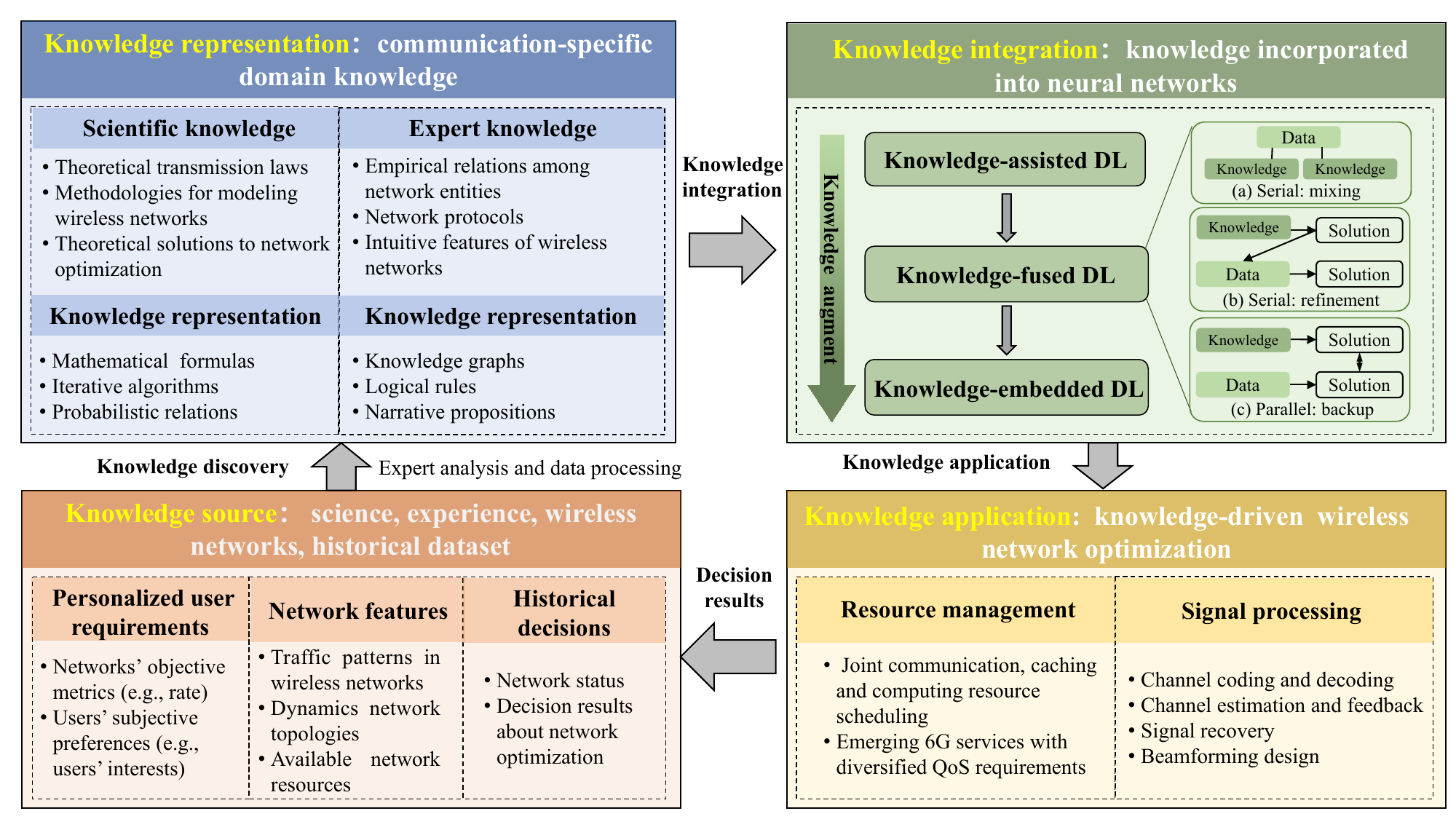}
\caption{\textcolor{black}{The holistic framework of knowledge-driven DL for wireless network optimization. Within this framework, knowledge source in wireless networks, communication-specific domain knowledge with various representations, knowledge integration approaches in neural networks and knowledge application in wireless networks are formed as a closed loop to continuously refine the latest knowledge and drive future wireless network optimization.}}
\label{Fig.1}
\end{figure*}

Knowledge-driven DL methods are appealing for network optimization as they possess the advantages of both model-based and data-driven methods.  Firstly, the interpretability of neural networks is significantly enhanced with domain knowledge integration, contributing to a better understanding and analysis of  the network's behavior. This interpretability also leads to lightweight knowledge-driven neural networks with reduced learnable parameters, decreased training data samples as well as relieved offline computational burden. Secondly, in contrast to  model-based theoretical methods depending on accurate modeling, domain knowledge used in knowledge-driven DL methods only offers a basic and general characteristic of the task, and the performance degradation caused by the model inaccuracy is compensated by the powerful universal approximation capability of DL.  Thirdly, knowledge-driven DL methods maintain the fast online inference ability of neural networks, which are able to meet the stringent low-latency requirements of real-time services in  dynamic networks. Owing to these strengths listed above, knowledge-driven DL  has the potential to efficiently and intelligently tackle complicated large-scale network optimization problems in 6G networks and will become a novel paradigm.

\section{The Framework of Knowledge-Driven DL in Wireless Networks}
In this section, to gain a holistic overview of knowledge-driven DL for network optimization,  we propose a framework concentrating on how knowledge circulates in wireless networks. As illustrated in Fig. 2, the proposed framework is constituted by knowledge source, knowledge representation, knowledge integration and knowledge application, which  form a closed loop and are respectively introduced in the sequel.

\subsection{Knowledge Source in Wireless Networks}

\textcolor{black}{Knowledge source indicates where to acquire knowledge in wireless networks.} In this field, domain knowledge usually stems from practical communication networks and is discovered by experts via logical analysis or data processing when exploring and understanding the principles of how communication networks work. For on-demand network optimization in 6G, in addition to existing communication-specific knowledge garnered over decades, new insights abstracted from personalized user requirements, network features and historical decisions are of critical importance. To provide customized services, the primary thing is characterizing personalized user requirements, which involves not only networks' objective metrics, such as rate and latency, but also users' subjective preferences analyzed by user behavior, including users' interests and moods. Then, capturing network features facilitates the effective network optimization design, which consists of identifying traffic patterns in wireless networks, sensing  dynamic network typologies and monitoring available network resources. Furthermore, drawing insights from historical decisions including network status and decision results about resource allocation and signal processing helps networks evolve and refine their optimization approaches, which contributes to the overall operational efficiency of wireless networks.

\subsection{ Knowledge Representation in Wireless Networks}

\textcolor{black}{After validation, experiences, insights and theories discovered in wireless networks are recorded in appropriate representations,  constituting communication-specific domain knowledge}, which can be classified into scientific knowledge and expert knowledge. Scientific knowledge aims to establish universal laws that can theoretically explain the operation mechanism of wireless networks. It is systematic and objective, and widely accepted by the communication community. Typically, scientific knowledge is represented in structured and formal mathematical expressions, including formulas, algorithms and probabilistic relations. Three representative examples of scientific knowledge in wireless networks are shown as follows. Theoretical transmission laws, such as  Shannon’s three theorems represented in formulas, lay the foundation principles of wireless communications. Methodologies for modeling wireless networks, such as the M/M/1 network queue model expressed in probabilistic relationships, offer insights into network behaviors. Theoretical solutions to network optimization, like the water-filling power allocation formed in iterative algorithms, provide upper-bound performance for network resource optimization.

Expert knowledge refers to knowledge that a group of experts acquire through extensive practice, learning and understanding of wireless networks. It is contextual and subjective, and  often specific to an individual's experiences, perspectives and interpretations. Although without theoretical verification, expert knowledge is highly valuable for solving complex real-world problems in wireless networks that require nuanced understanding and decision-making. Compared with scientific knowledge, expert knowledge is not easy to articulate and usually represented in relatively informal expressions. Some examples of expert knowledge in wireless networks are outlined below.  Empirical relations among network entities are extracted in the form of knowledge graphs for scenario recognition or network operation and management. Network protocols in real-world networks written in logical rules are often designed based on accumulated experiences. Intuitive features of wireless networks represented by narrative propositions, such as the temporal correlation of network traffic,  are abstracted to facilitate wireless network optimization.   

\subsection{Knowledge Integration in Wireless Networks}
\textcolor{black}{Communication-specific domain knowledge with various representations is incorporated into neural networks via knowledge integration.} As illustrated in Fig.~2, there are three main knowledge integration approaches to designing knowledge-driven DL, i.e., knowledge-assisted, knowledge-fused and knowledge-embedded DL. In knowledge-assisted DL, domain knowledge is leveraged to guide the selection of appropriate neural networks. Taking a step forward, knowledge-fused DL operates by integrating model-based theoretical methods with data-driven neural networks to respectively address decoupled sub-problems of a complex problem. Moving further, knowledge-embedded DL deeply embeds domain knowledge into restructured neural network models. It is worth pointing out that, from knowledge-assisted neural networks through knowledge-fused neural networks to knowledge-embedded neural networks, domain knowledge plays an increasingly pivotal role. \textcolor{black}{A comprehensive explanation of the techniques and applications for each knowledge integration approach is provided in Section \uppercase\expandafter{\romannumeral4}.}


\subsection{Knowledge Application in Wireless Networks}
\textcolor{black}{Designed knowledge-driven DL methods are applied to diversified fields in wireless networks, including resource management and signal processing.} For each application, appropriate knowledge integration approaches should be carefully selected or combined to boost the network performance. In resource management, to support emerging 6G services with diversified quality of service (QoS) requirements, both scientific and expert knowledge is incorporated into neural networks to  effectively address large-scale complicated resource orchestration tasks, including  joint communication, caching and computing resource scheduling, which relieves the online computational burden and  accurate modeling pressure. In signal processing, knowledge-driven DL is adopted in  channel coding and decoding, channel state information estimation and feedback, signal recovery and beamforming design.  For these tasks, a wealth of theoretical methods has been accumulated over decades, which is utilized to  inform the design of customized neural networks, thereby enhancing the performance and accelerating the response time.

\textcolor{black}{Once network optimization has taken place, the network controller archives network status and decision results as historical data, which is then evaluated and abstracted as domain knowledge to optimize wireless networks.}  By maintaining this cycle of knowledge discovery, knowledge integration and knowledge application, the knowledge loop continuously refines the latest knowledge and drives future wireless network optimization, ultimately realizing intelligent native 6G networks.

\section{The Taxonomy of Knowledge Integration Approaches in Wireless Networks}
In this section, three main knowledge integration approaches, i.e.,  knowledge-assisted, knowledge-fused and knowledge-embedded DL, are presented in detail. For each category, we describe its primary integrated knowledge, knowledge representations and potential applications in wireless networks, as listed in Fig.~3.

\begin{figure*}[htbp]
\centering
\includegraphics[width=0.98\textwidth, height=0.6\textheight]{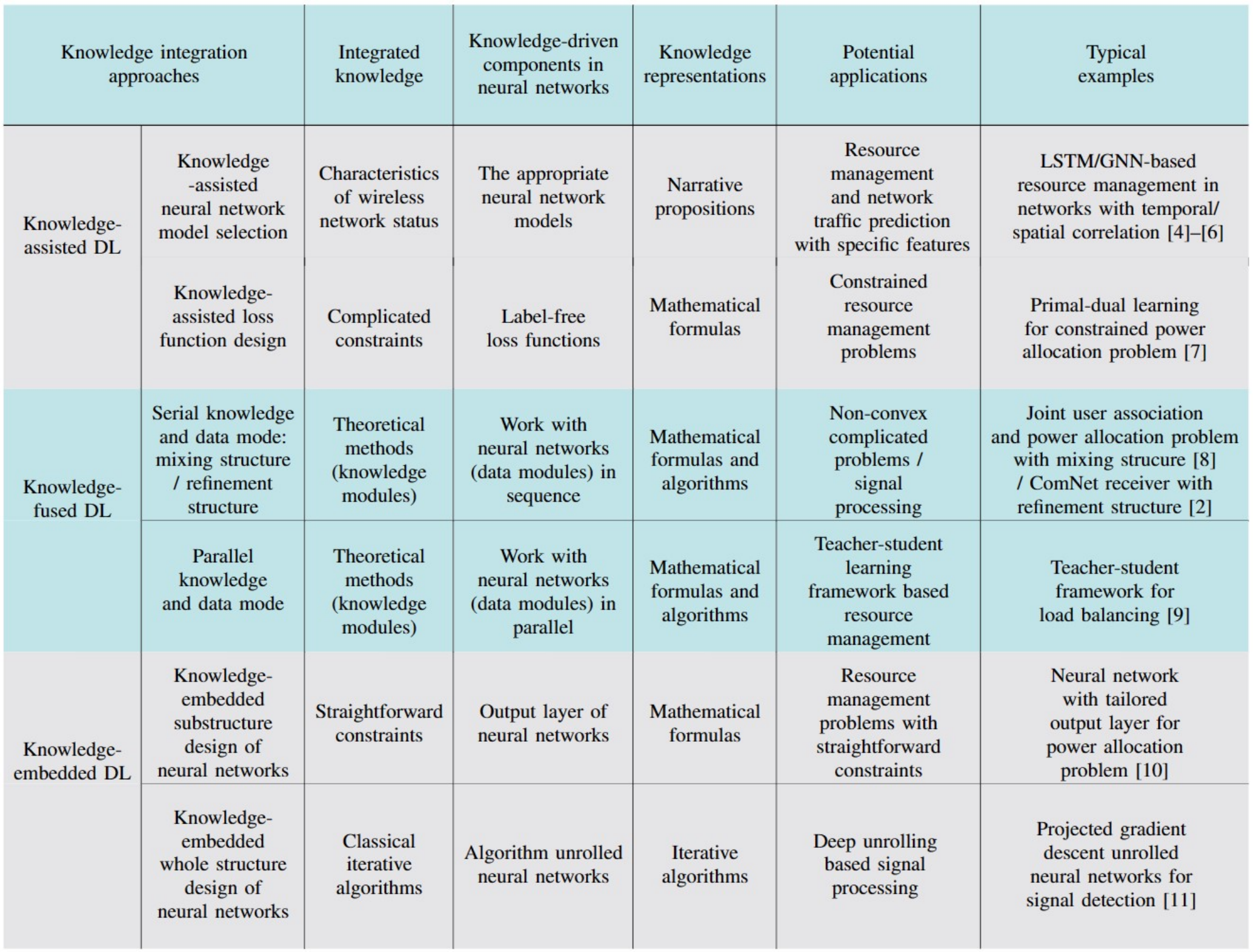}
\caption{\textcolor{black}{Three Main Knowledge Integration Approaches and their Corresponding Integrated Knowledge, Knowledge-driven Components in Neural Networks, Knowledge Representations, Potential Applications and Typical Examples in Wireless Networks.}}
\label{Fig.3}
\end{figure*}

\subsection{ Knowledge-Assisted DL in Wireless Networks}
Knowledge-assisted DL involves prior knowledge to guide the choice of  neural network models and the determination of loss functions. In wireless networks, neural network models  are usually led by prior knowledge concerning the properties of tasks, which  stems from an in-depth understanding and experiential recognition of tasks and is represented as narrative propositions. Typically, two  characteristics  of network status, i.e., the temporal correlation and the spatial correlation, respectively make long short-term memory (LSTM) and graph neural network (GNN) become appealing neural network models in wireless networks. LSTM, with its proficiency in learning temporal dependencies within sequence data via cell states and control gates, is widely utilized for long-term network resource allocation tasks, such as minimizing age of information \cite{leng2022learning}, and network traffic prediction \cite{hua2019deep}.  By incorporating the spatial graph-structured network topology into neural network models, GNN exhibits enhanced scalability and can effectively manage dynamic wireless tasks, which have been employed for tasks such as power allocation, link scheduling and network slicing  in dynamic networks \cite{he2021overview}.

Knowledge-assisted loss function involves domain knowledge, typically in the form of mathematical formulas derived from scientific knowledge, as label-free terms in loss functions to foster neural networks learn from both data and knowledge. In wireless networks, a widely adopted knowledge-assisted loss function is the constraint-specific loss function, which additionally adds constraints of network optimization problems as penalty terms in loss functions to penalize any violations of the constraints.  A representative learning approach employing  the constraint-specific loss is primal-dual learning \cite{eisen2019dual}, which introduces non-negative dual variables and converts constraints of  problems  as additional linear penalty terms in the objective function via Lagrangian duality techniques.  This primal-dual learning has seen successful applications in constrained resource management problems in wireless networks, such as probabilistic constrained power allocation problems for video streaming transmission, joint power and bandwidth allocation for ultra-reliable and low-latency communications, and power allocation problem with statistical constraints in ad-hoc networks \cite{liu2020optimizing}.

\subsection{Knowledge-Fused DL in Wireless Networks}
Knowledge-fused DL leverages the combination of model-based theoretical methods and data-driven neural networks to address complex wireless network problems, by dividing the original problems into multiple interrelated sub-problems, each addressed by either theoretical methods (i.e., knowledge modules) or neural networks (i.e., data modules) hinged upon their unique attributes. Typically, sub-problems that have optimal analytical solutions or complex constraints are managed with knowledge modules, while those with high-dimensional, nonlinear, and challenging-to-model mapping relationships are addressed by data modules. The integrated knowledge here primarily stems from scientific knowledge in the forms of mathematical formulas and algorithms. In knowledge-fused DL, knowledge and data modules can operate in serial or parallel modes, as illustrated in Fig. 2.

For the serial configuration, knowledge and data modules work sequentially to collectively tackle a specific problem. According to whether knowledge and data modules address the same sub-problem,  the serial mode consists of mixing  and refinement structures. For mixing structures, knowledge and data modules alternatively address different sub-problems,  playing to their respective strengths. Owing to this benefit,  mixing structures have been applied to efficiently solve complicated non-convex resource management problems  in wireless networks, such as the joint user association and power allocation problems in cell-free systems \cite{guenach2022deep}.   
For refinement structures,  knowledge modules initially provide rough solutions based on generalized theoretical models, and the following data module refines these solutions by exploiting real-world  data. Thanks to the initial solutions generated by knowledge modules, the number of training data samples required for data modules is  substantially reduced. In wireless networks, refinement structures are  widely applied in signal processing  with a wealth of theoretical algorithms. An exemplary instance  is the ComNet receiver proposed in \cite{he2019modeldriven}, where both channel estimation and signal detection use existing communication algorithms for initialization, followed by neural networks to refine coarse solutions. 

In the parallel mode of knowledge-fused neural networks, knowledge and data modules operate concurrently and independently to address the same problem, with the final output being an effective combination  of them following certain rules. This mode enhances the robustness and reliability of the overall system by creating a backup system, where each module can offset the potential deficiencies of the other. A typical example  in wireless  networks is the teacher-student learning framework proposed in \cite{zheng2021leveraging}, where a teacher module represented by explainable theoretical algorithms and a student block represented by dynamic RL concurrently solve the targeted problem.  The superior output from the teacher module is absorbed by the student neural network through reward shaping to improve the robustness. This teacher-student framework has been applied in loading balancing and TCP congestion control problems and exhibits more stable performance in dynamic environments.

\begin{figure*}[htbp]
\centering
\includegraphics[width=0.98\textwidth, height=0.46\textheight]{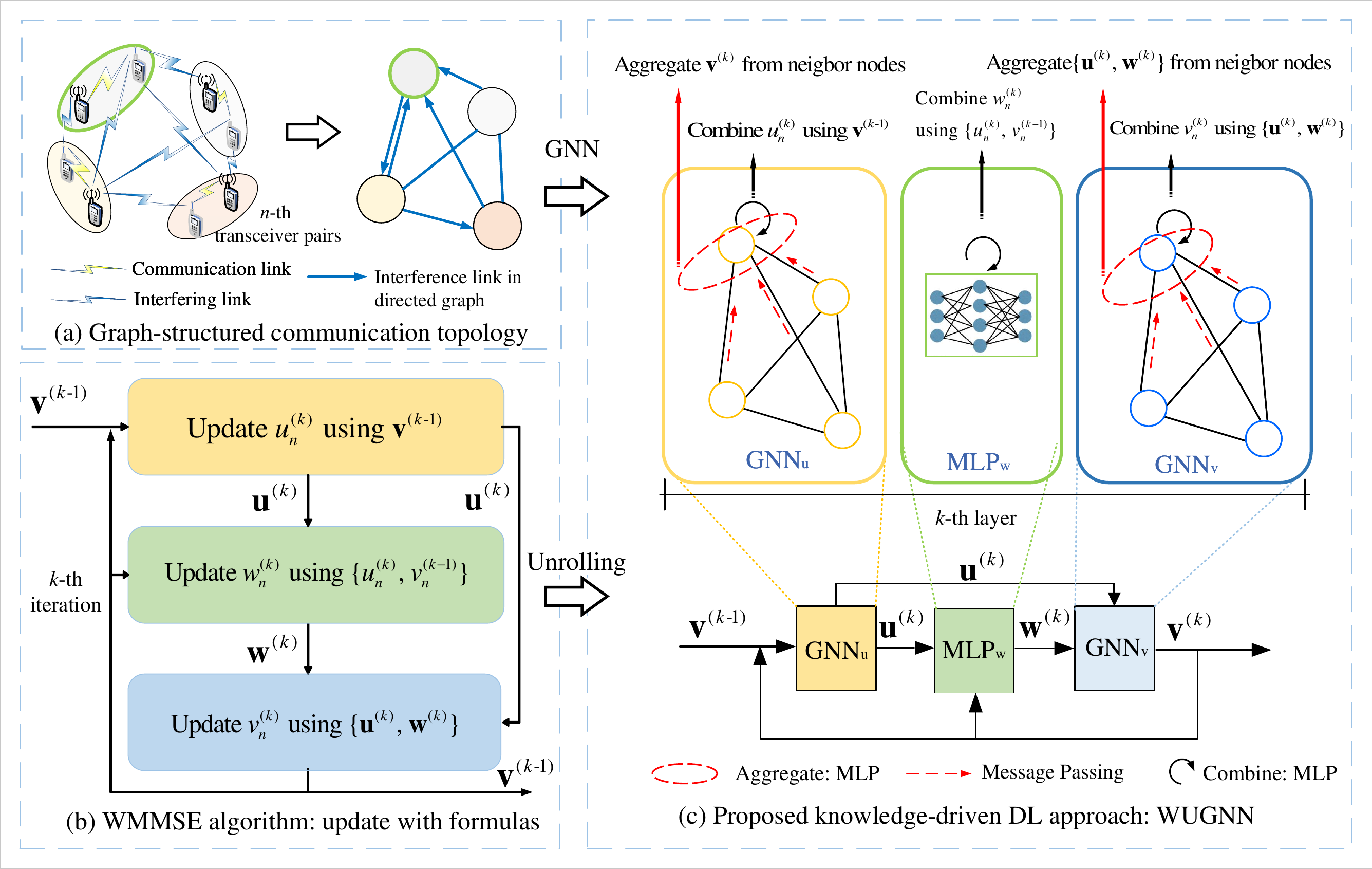}
\caption{\textcolor{black}{The proposed knowledge-driven DL approach for resource management in D2D networks. Both the graph-structured communication topology in D2D networks and the classical WMMSE algorithm for the sum rate maximization problem are integrated into neural networks, inspiring the proposed WMMSE algorithm unrolled GNN. In this figure, $\textbf{u}$, $\textbf{w}$ and $\textbf{v}$ are three variable blocks in the classical WMMSE algorithm and their scalar forms are variables of the $n$-th transceiver pair in D2D networks.}}
\label{Fig.1}
\end{figure*}

\subsection{Knowledge-Embedded DL in Wireless Networks}
In knowledge-embedded DL, domain knowledge is internally embedded into the customized neural network models, enhancing interpretability. Rather than adopting generic neural network models, the structure of knowledge-embedded neural networks is tailored to the target problem, guided by communication-specific knowledge. In wireless networks, such embedded knowledge usually comes from explainable scientific principles, in the form of mathematical formulas and algorithms. Knowledge-embedded DL includes knowledge-embedded substructure design and knowledge-embedded whole structure design in neural networks. 

Knowledge-embedded substructure design of neural networks is embedding domain knowledge into the local structure of neural networks. Typically, prior knowledge is integrated into the output layer of neural networks to guarantee that the networks' final output is within the feasible region of the targeted problem. This method has seen early adoption in wireless resource management problems that have relatively straightforward constraints. For instance, a  customized neural network with a knowledge-embedded substructure was presented in \cite{li2021multicell} to tackle the power allocation problem in downlink multicell scenarios aiming for the sum rate maximization. Specifically, to ensure the per-user rate constraints of the problem, a knowledge-embedded output layer based on a geometrical interpretation of these constraints is appended after a deep neural network to project the power allocation results onto the feasible domain.  Benefiting from this embedded knowledge layer, the proposed novel neural network can effectively manage the model mismatch problem between training and testing datasets.

Knowledge-embedded whole structure design of neural networks builds neural network models using domain-specific knowledge. The most representative approach is deep unrolling, converting an iterative algorithm into a deep neural network by "unrolling" the iterations of the algorithm into the layers of the neural network \cite{monga2021algorithm}.  The resulting unrolled neural network mirrors the structure and logic of the original algorithm, and enables the fixed parameters in the algorithm to be flexibly learnable parameters. By bringing together the structural insights of proven iterative algorithms with the learning capabilities of neural networks, deep unrolling improves the  performance of conventional algorithms while boosting the interpretability of neural networks.  Hence, deep unrolling  has been widely adopted across a range of intelligent signal processing tasks, such as signal detection, beamforming design, channel estimation and user activity detection. Classic iterative signal processing algorithms like projected gradient descent, approximate message passing, alternating direction method of multipliers, and weighted minimum mean square error (WMMSE) have been unrolled into neural networks to enhance the system performance \cite{jagannath2021redefining}.


\section{A Case Study: WMMSE Algorithm Unrolled GNN for Resource Management}

In this section, we provide a case study illustrating the effectiveness of knowledge-driven DL for resource management in wireless networks. Consider a device-to-device (D2D) scenario with $N$ transceivers,  as depicted in Fig. 4(a), the task is to coordinate interference through power allocation to maximize the sum rate. Typically, this resource management problem is addressed by the model-based iterative WMMSE algorithm \cite{shi2011iteratively}, as shown in Fig. 4(b). While the WMMSE algorithm can achieve the Karush-Kuhn-Tucker point of the sum rate maximization problem by alternately updating three variable blocks ${\textbf{u}}$, ${\textbf{w}}$ and ${\textbf{v}}$ with mathematical formulas, the solution it finds is a local optimum, not the global one. Moreover, the WMMSE algorithm needs to be executed in real time whenever network parameters change, and the growing scale of the network increases the computational complexity, leading to unacceptable online processing time.

To overcome these challenges, this study proposes a novel knowledge-driven DL approach for intelligent resource management, which integrates both the graph-structured network topology inherent in the D2D scenario and the model-based WMMSE algorithm. 
In specific, the D2D network topology naturally forms a graph where each transceiver pair is a node and each interference link is an edge. This insight informs  GNN as the basic neural network model. Besides, to further enhance interpretability, the whole structure of neural networks is constructed by unrolling the classical WMMSE algorithm. \textcolor{black}{The resulting framework, i.e., the WMMSE algorithm unrolled GNN (WUGNN), is depicted in Fig. 4(c), which is a $K$-layer neural network that mimics the WMMSE algorithm with $K$ iterations. Each layer includes two GNN modules, i.e., $\text{GNN}_u$ and $\text{GNN}_v$, to respectively update  ${\textbf{u}}$ and ${\textbf{v}}$ with information aggregation from neighboring nodes as well as one multilayer perceptron (MLP) module $\text{MLP}_w$ to update ${\textbf{w}}$ with node $n$'s own information. Both aggregation and combine functions in $\text{GNN}_u$ and $\text{GNN}_v$ are also MLPs to learn the corresponding update formulas in the WMMSE algorithm.  This structure within each layer of the WUGNN is inspired by the updates of ${\textbf{u}}$, ${\textbf{w}}$ and ${\textbf{v}}$ in the WMMSE algorithm.  
As it involves the network topology to select GNN as the basic model and unrolls iterations in the WMMSE algorithm as layers in neural networks, the proposed WUGNN simultaneously belongs to knowledge-assisted neural network model selection and knowledge-embedded whole structure design of neural networks. }

\begin{figure}[htbp]
\centering
\includegraphics[width=0.45\textwidth]{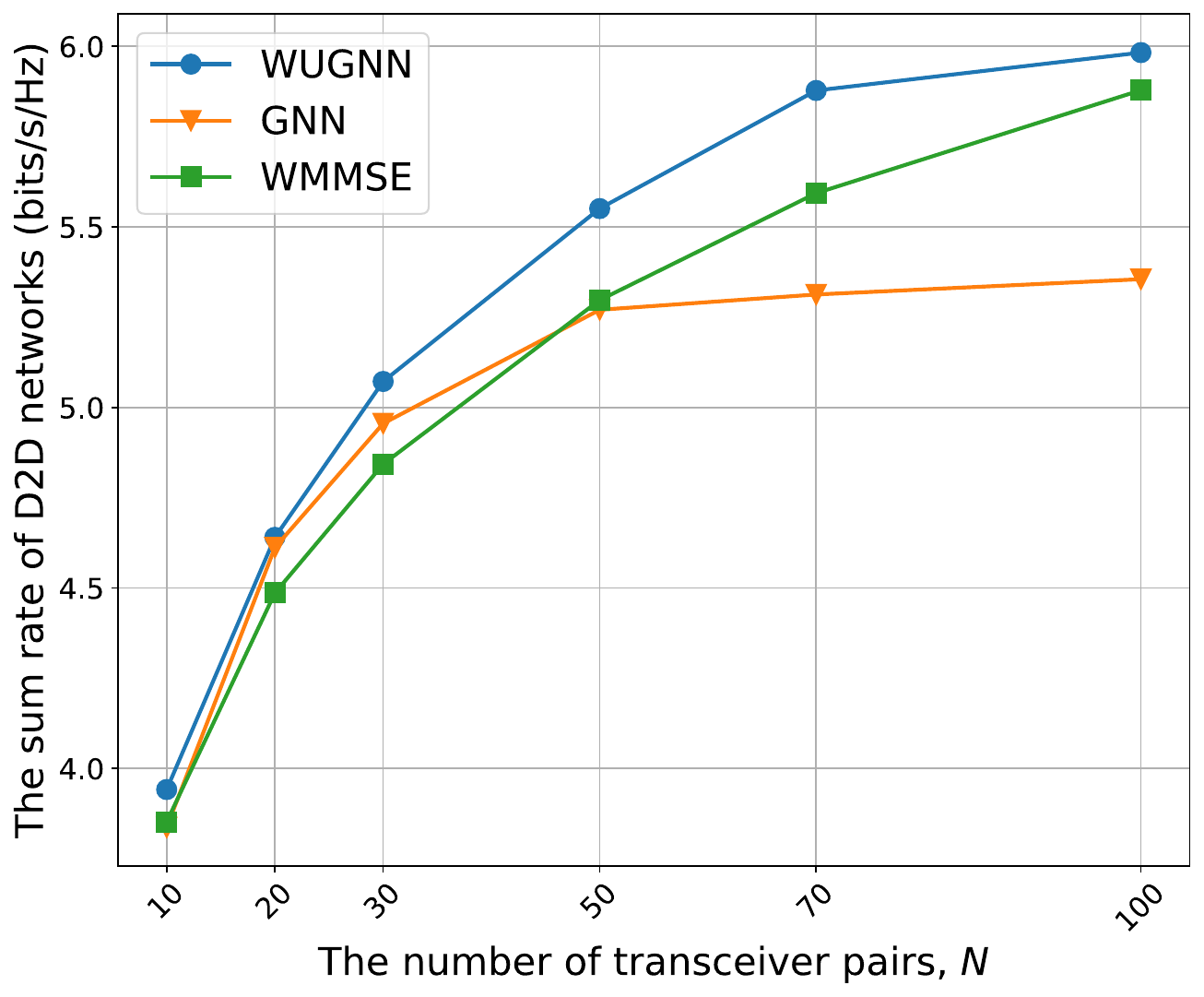}
\caption{The scalability comparison of the proposed WUGNN and other two approaches.}
\label{Fig:simulation}
\end{figure}

In the simulation, the WMMSE algorithm and GNN approach \cite{shen2021graph} are also conducted for comparison. Neural network-related approaches, i.e., GNN and the proposed WUGNN, are initially trained in a scenario with 10 transceiver pairs and subsequently applied to new scenarios with a varying number of pairs. The model-based WMMSE algorithm is re-executed for each new scenario and serves as a performance benchmark. As depicted in Fig. \ref{Fig:simulation}, our proposed knowledge-driven WUGNN outperforms the other two approaches in terms of the sum rate as the network size expands from 10 to 100, demonstrating strong scalability. By contrast, the sum rate of the GNN approach only surpasses that of the WMMSE algorithm when the network size is less than 50. Furthermore, Table \uppercase\expandafter{\romannumeral1} presents the online inference time required by these three approaches, which is counted on a server with 11th Gen Intel(R) Core(TM) i7-11700 @ 2.50GHz, GPU: NVIDIA GeForce RTX 3060. As expected, the online time consumption of two neural network-related approaches, including our proposed solution, falls in the millisecond level and  is nearly a hundred times less than that of the WMMSE algorithm. Therefore, by taking into account the sum rate performance, the scalability, and the online processing time, our proposed WUGNN is demonstrated to be the most efficient solution for the considered resource management task, which can swiftly adapt to real-time changes in large-scale dynamic networks while assuring the network performance.

\begin{table}[!hbt]
    \centering
    \renewcommand\arraystretch{1.5}
    \caption{The Online Inference Time of Three Approaches (Unit: Seconds).}
    \resizebox{1.0\linewidth}{!}{
    \begin{tabular}{ccccccc}
    \hline
    \cellcolor[HTML]{c1e4e9}$N$ &\cellcolor[HTML]{e5e4e6}10 & \cellcolor[HTML]{c1e4e9}20  & \cellcolor[HTML]{e5e4e6} 30& \cellcolor[HTML]{c1e4e9}50 & \cellcolor[HTML]{e5e4e6}70&\cellcolor[HTML]{c1e4e9}100  \\
    \hline\hline
    \cellcolor[HTML]{c1e4e9}WUGNN &\cellcolor[HTML]{e5e4e6}0.018 & \cellcolor[HTML]{c1e4e9}0.021  & \cellcolor[HTML]{e5e4e6} 0.022& \cellcolor[HTML]{c1e4e9}0.021 & \cellcolor[HTML]{e5e4e6}0.025&\cellcolor[HTML]{c1e4e9}0.040  \\
    \hline
    \cellcolor[HTML]{c1e4e9}GNN &\cellcolor[HTML]{e5e4e6}0.008 & \cellcolor[HTML]{c1e4e9}0.010  & \cellcolor[HTML]{e5e4e6} 0.010& \cellcolor[HTML]{c1e4e9}0.009 & \cellcolor[HTML]{e5e4e6}0.016&\cellcolor[HTML]{c1e4e9}0.021  \\
    \hline
    \cellcolor[HTML]{c1e4e9}WMMSE &\cellcolor[HTML]{e5e4e6}4.262 & \cellcolor[HTML]{c1e4e9}20.239  & \cellcolor[HTML]{e5e4e6} 41.328& \cellcolor[HTML]{c1e4e9}114.669 & \cellcolor[HTML]{e5e4e6}188.005&\cellcolor[HTML]{c1e4e9}384.590  \\
    \hline
    \end{tabular}
    }
\end{table}

\section{Open Issues}
Although simultaneously preserving the interpretability of domain knowledge, as well as the powerful universal approximation ability and the fast online inference of neural networks, knowledge-driven DL for wireless network optimization is still in its infancy and faces several open issues, which are discussed in the following.

\subsection{Knowledge-Driven DL Handling Complex Constraints} 

While significant advancements have been made in knowledge-driven DL for wireless network optimization, they only well address  problems with relatively straightforward constraints, such as linear functions. Such simple constraints can be translated into additional blocks following neural networks, which project the outputs into feasible regions of optimization problems. Nevertheless, in the era of 5G beyond and 6G,  newly emerging applications are characterized by diversified QoS requirements, and novel air-space-ground-sea integrated communication networks are featured with high-dynamic communication typologies and multi-dimensional heterogeneous resources. These factors lead to resource management problems with complex non-convex constraints, in the forms of quadratic, logarithmic, fractional functions, and so on. Although those problems can be tackled by  constraint-specific loss functions to a certain degree, they introduce extra penalty factors that need to be predetermined and fail to strictly enforce the outputs to comply with the constraints.  Therefore, developing reliable knowledge-driven DL that ensures the outputs strictly satisfy  any nonlinear constraint is an opening and challenging issue,  particularly in the realm of wireless resource management with stringent QoS constraints. 

\subsection{Theoretical Analysis of Knowledge-Driven DL}
Despite the impressive performance exhibited by recent data-driven DL methods, their adoption for  practical wireless network optimization is hampered by the lack of solid theoretical foundations. Integrating communication-specific domain knowledge into neural networks can enhance the interpretability of neural networks, as knowledge-driven DL is typically inspired by explainable models with rigorously analytical results. In particular, the deep unrolling approach constructs customized neural networks by inheriting the structure and operations  of iterative algorithms, which is  especially promising.  However, thoroughly understanding  the mechanism of knowledge-driven DL remains elusive. Moreover, it is difficult for knowledge-driven DL  to provide  robust performance guarantees as model-based theoretical methods do. Therefore, to apply knowledge-driven DL in wireless networks with highly reliable QoS, such as industrial IoT scenarios, more work needs to be done in future research to carry out a  profound theoretical analysis of knowledge-driven DL. This cognition will offer a level of transparency that's highly desirable when dealing with complex network optimization problems and provide an avenue to identify areas for further optimization.

\subsection{ Knowledge Selection and Aggregation  for Integration}
To design a knowledge-driven DL for a specific task, the initial step is identifying the suitable domain knowledge to incorporate, which  currently  relies on experts' handcrafted empirical cognition.  As the same task may have several different sides and experts also have varied understandings and insights, different pieces of knowledge for the same task are acquired, consequently leading to divergent knowledge-driven DL designs.
For instance, when addressing a power allocation task in wireless networks, incorporating knowledge about the graph-structured communication topology results in a  GNN-based DL, while employing knowledge about the classical  WMMSE algorithm leads to a WMMSE-unrolled DL. Hence, selecting and aggregating critical domain knowledge into neural networks to achieve the best performance is an open issue. Furthermore, network optimization in future 6G networks is a multifaceted endeavor, necessitating synergistic contributions from experts across various fields. For example, the comprehensive characterization of the individual user's QoS needs for diversified services requires consideration of both subjective factors informed by psychological analyses and objective service metrics derived from network engineering. How to aggregate such experts' insights from different fields,  potentially with the aid of tools like knowledge graphs or ChatGPT, to facilitate knowledge-driven intelligent network optimization merits further exploration in future research. 



\section{Conclusions}
This article has given a holistic overview of knowledge-driven DL for wireless network optimization. First, the concept and strengths of knowledge-driven DL have been presented. Then, a framework indicating the knowledge circulation process in wireless networks has been proposed.  What is more, knowledge integration approaches consisting of knowledge-assisted, knowledge-fused and knowledge-embedded DL have been highlighted. In addition, a case study about intelligent resource management has  been investigated, demonstrating the scalability, competitive performance and fast inference time of knowledge-driven DL.




\ifCLASSOPTIONcaptionsoff
  \newpage
\fi



%

\section*{Acknowledgements}
This work was supported by the National Key Research and Development Program of China under Grant 2020YFB1807700.

\bibliographystyle{IEEEtran}

\bibliography{references}

\begin{thebibliography}{10}
\providecommand{\url}[1]{#1}
\csname url@samestyle\endcsname
\providecommand{\newblock}{\relax}
\providecommand{\bibinfo}[2]{#2}
\providecommand{\BIBentrySTDinterwordspacing}{\spaceskip=0pt\relax}
\providecommand{\BIBentryALTinterwordstretchfactor}{4}
\providecommand{\BIBentryALTinterwordspacing}{\spaceskip=\fontdimen2\font plus
\BIBentryALTinterwordstretchfactor\fontdimen3\font minus \fontdimen4\font\relax}
\providecommand{\BIBforeignlanguage}[2]{{%
\expandafter\ifx\csname l@#1\endcsname\relax
\typeout{** WARNING: IEEEtran.bst: No hyphenation pattern has been}%
\typeout{** loaded for the language `#1'. Using the pattern for}%
\typeout{** the default language instead.}%
\else
\language=\csname l@#1\endcsname
\fi
#2}}
\providecommand{\BIBdecl}{\relax}
\BIBdecl

\bibitem{ITU2023framework}
ITU-R, ``Framework and overall objectives of the future development of {IMT} for 2030 and beyond,'' \emph{Draft New Recommendation}, June 2023, available online: https://www.itu.int/en/ITU-R/study-groups/rsg5/rwp5d/imt-2030/Pages/default.aspx.

\bibitem{he2019modeldriven}
H.~He, S.~Jin, C.-K. Wen, F.~Gao, G.~Y. Li, and Z.~Xu, ``Model-driven deep learning for physical layer communications,'' \emph{IEEE Wireless Commun.}, vol.~26, no.~5, pp. 77--83, Oct. 2019.

\bibitem{shlezinger2021model}
N.~Shlezinger, J.~Whang, Y.~C. Eldar, and A.~G. Dimakis, ``Model-based deep learning: Key approaches and design guidelines,'' in \emph{Proc. IEEE Data Science and Learning Workshop (DSLW)}, June 2021, pp. 1--6.

\bibitem{leng2022learning}
S.~Leng and A.~Yener, ``Learning to transmit fresh information in energy harvesting networks,'' \emph{IEEE Trans. Green Commun. and Netw.}, vol.~6, no.~4, pp. 2032--2042, Dec. 2022.

\bibitem{hua2019deep}
Y.~Hua, Z.~Zhao, R.~Li, X.~Chen, Z.~Liu, and H.~Zhang, ``Deep learning with long short-term memory for time series prediction,'' \emph{IEEE Commun. Mag.}, vol.~57, no.~6, pp. 114--119, June 2019.

\bibitem{he2021overview}
S.~He, S.~Xiong, Y.~Ou, J.~Zhang, J.~Wang, Y.~Huang, and Y.~Zhang, ``An overview on the application of graph neural networks in wireless networks,'' \emph{IEEE Open J. Commun. Soc.}, vol.~2, pp. 2547--2565, 2021.

\bibitem{eisen2019dual}
M.~Eisen, C.~Zhang, L.~F.~O. Chamon, D.~D. Lee, and A.~Ribeiro, ``Dual domain learning of optimal resource allocations in wireless systems,'' in \emph{Proc. IEEE International Conference on Acoustics, Speech and Signal Processing (ICASSP)}, May 2019, pp. 4729--4733.

\bibitem{liu2020optimizing}
D.~Liu, C.~Sun, C.~Yang, and L.~Hanzo, ``Optimizing wireless systems using unsupervised and reinforced-unsupervised deep learning,'' \emph{IEEE Network}, vol.~34, no.~4, pp. 270--277, July 2020.

\bibitem{guenach2022deep}
M.~Guenach, A.~A. Gorji, and A.~Bourdoux, ``A deep neural architecture for real-time access point scheduling in uplink cell-free massive {MIMO},'' \emph{IEEE Trans. Wireless Commun.}, vol.~21, no.~3, pp. 1529--1541, Mar. 2022.

\bibitem{zheng2021leveraging}
Y.~Zheng, ``Leveraging domain knowledge for robust deep reinforcement learning in networking,'' \emph{Proc. IEEE International Conference on Computer Communications (INFOCOM)}, 2021.

\bibitem{li2021multicell}
Y.~Li, S.~Han, and C.~Yang, ``Multicell power control under rate constraints with deep learning,'' \emph{IEEE Trans. Wireless Commun.}, vol.~20, no.~12, pp. 7813--7825, Dec. 2021.

\bibitem{monga2021algorithm}
V.~Monga, Y.~Li, and Y.~C. Eldar, ``Algorithm unrolling: Interpretable, efficient deep learning for signal and image processing,'' \emph{IEEE Signal Processing Mag.}, vol.~38, no.~2, pp. 18--44, Mar. 2021.

\bibitem{jagannath2021redefining}
A.~Jagannath, J.~Jagannath, and T.~Melodia, ``Redefining wireless communication for {6G}: Signal processing meets deep learning with deep unfolding,'' \emph{IEEE Trans. Artif. Intell.}, vol.~2, no.~6, pp. 528--536, Dec. 2021.

\bibitem{shi2011iteratively}
Q.~Shi, M.~Razaviyayn, Z.-Q. Luo, and C.~He, ``An iteratively weighted {MMSE} approach to distributed sum-utility maximization for a {MIMO} interfering broadcast channel,'' \emph{IEEE Trans. Signal Processing}, vol.~59, no.~9, pp. 4331--4340, Apr. 2011.

\bibitem{shen2021graph}
Y.~Shen, Y.~Shi, J.~Zhang, and K.~B. Letaief, ``Graph neural networks for scalable radio resource management: Architecture design and theoretical analysis,'' \emph{IEEE J. Sel. Areas Commun.}, vol.~39, no.~1, pp. 101--115, Jan. 2021.

\end{thebibliography}

%

%
\begin{IEEEbiographynophoto}{Ruijin Sun}
 received the Ph.D. degree from the Beijing University of Posts and Telecommunications, China, in 2019. She is currently a Lecturer with the State Key Lab of ISN and the School of Telecommunications Engineering, Xidian University, Shanxi, China. She worked as a joint Postdoctoral Fellow with Peng Cheng Laboratory and Tsinghua University from 2019 to 2021. Her research interests are in the area of knowledge-driven wireless resource allocation and MIMO ad-hoc networks.
\end{IEEEbiographynophoto}

\begin{IEEEbiographynophoto}{Nan Cheng}
 received the Ph.D. degree from the Department of Electrical and Computer Engineering, University of Waterloo in 2016, and B.E. degree and the M.S. degree from the Department of Electronics and Information Engineering, Tongji University, Shanghai, China, in 2009 and 2012, respectively. He worked as a Post-doctoral fellow with the Department of Electrical and Computer Engineering, University of Toronto, from 2017 to 2019. He is currently a professor with State Key Lab. of ISN and with School of Telecommunications Engineering, Xidian University, Shaanxi, China. His current research focuses on B5G/6G, space-air-ground integrated network, and self-driving system. His research interests also include AI-driven future networks.
\end{IEEEbiographynophoto}

\begin{IEEEbiographynophoto}{Changle Li}
 received the Ph.D. degree in communication and information system from Xidian University, China, in 2005. He conducted his post-doctoral research in Canada and the National Institute of Information and Communications Technology, Japan, respectively. He had been a Visiting Scholar with the University of Technology Sydney. He is currently a Professor with the State Key Laboratory of ISN, Xidian University. His research interests include intelligent transportation systems, vehicular networks, mobile ad hoc networks, and wireless sensor networks.
\end{IEEEbiographynophoto}

\begin{IEEEbiographynophoto}{Fangjiong Chen}
 received a B.S. degree in electronics and information technology from Zhejiang University, Hangzhou, China, in 1997 and received a Ph.D. degree in communication and information engineering from South China University of Technology, Guangzhou, China, in 2002. He was a senior research associate in the Department of Computer Science, City University of Hongkong, Hongkong, China, from July 2006 to June 2007. Currently, he is a full-time Professor at the School of Electronics and Information Engineering, South China University of Technology. He is the Director of the Guangdong Provincial Key Laboratory of Short-Range Wireless Detection and Communication. Prof. Chen is a co-recipient of the Best Demo Award of ACM WUWNET2008 and the Best Paper Award of ICNC2016, ICCT2019, and UCOM2023. His research interests include signal detection and estimation, array signal processing, and wireless communication.
\end{IEEEbiographynophoto}

\begin{IEEEbiographynophoto}{Wen Chen}
 is currently a tenured Professor with the Department of Electronic Engineering, Shanghai Jiao Tong University, China, where he is also the Director of the Broadband Access Network Laboratory. He has published more than 100 articles in IEEE journals and more than 120 papers in IEEE conferences, with citations more than 10000 in Google scholar. His research interests include reconfigurable meta-surface, multiple access, wireless AI, and green networks.
\end{IEEEbiographynophoto}

%
%




\end{document}